\input phyzzx.tex

\tolerance=500000
\overfullrule=0pt
\pubnum={}
\date={}
\titlepage

\hskip6cm {BRX-TH-409, US-FT-13-97, BROWN-HET-1080}

\hskip11cm{hep-th/9704174}

\hskip11cm{April, 1997.}

\title{On the Picard--Fuchs Equations for Massive $N=2$ Seiberg--Witten
Theories}
\author{Jos\'e M. Isidro,$^a$ \foot{Research supported by the DOE under grant
DE-FG02-92ER40706. Home address: Dept. de F\'{\i}sica de Part\'{\i}culas, 
Universidad de Santiago, 15706 Santiago, Spain.}
Avijit Mukherjee,$^a$\foot{Research supported by
the DOE under grant DE-FG02-92ER40706.}
Jo\~ao~P.~Nunes,$^{b}$\foot{Research supported by  the DOE under
grant DE-FG0291ER40688-Task A. } and Howard J. Schnitzer$^a
$\foot{Research supported in part by
the DOE under grant DE-FG02-92ER40706.}
\foot{{\tt isidro,
mukherjee, schnitzer@binah.cc.brandeis.edu, nunes@het.brown.edu}}}

\address{Department of Physics, Brandeis University, Waltham MA
 02254-9100, USA.$^a$}

\address{ Department  of Physics, Brown University, Providence RI
 02912, USA.$^b$}

\abstract{ A new method to obtain the Picard--Fuchs equations of effective, 
$N=2$ supersymmetric
gauge theories with massive matter hypermultiplets in the fundamental 
representation is presented.
It generalises a previously described method to derive the Picard--Fuchs
 equations of both pure super
Yang--Mills and supersymmetric gauge theories with massless matter 
hypermultiplets. The techniques
developed are well suited to symbolic computer calculations. }

\endpage

\def\d{{\rm d}}
\def\p{\partial}
\def\o{\over}
\def\ie{{\it i.e.}}

\def\np{Nucl. Phys.}
\def\pl{Phys. Lett.}
\def\prl{Phys. Rev. Lett.}
\def\pr{Phys. Rev.}

\def\ijmp{Int. J. Mod. Phys.}

\REF\SW{N. Seiberg and E. Witten\journal\np&B426(94)19, {\tt (hep-th/9407087)}
\journal\np&B431(94)484 {\tt (hep-th/9408095)}.}

\REF\KLEMM{A. Klemm, W. Lerche, S. Theisen and S. Yanckielowiz
\pl {\bf B344} (1995) 169 {\tt (hep-th/9411048)}.}

\REF\ARG{P.C. Argyres  and A.E. Faraggi, \prl {\bf 74} (1995) 3931 
{\tt (hep-th/9411057)}.}

\REF\BRAND{A. Brandhuber and K. Landsteiner, \pl {\bf B358} (1995) 73  
{\tt (hep-th/9507008)}.}

\REF\SHAPERE{P.C. Argyres and A. Shapere, \np {\bf B461} (1996) 437
{\tt (hep-th/9509175)}.}

\REF\ABOLHASANI{M.R. Abolhasani, M. Alishahiha and A.M. Ghezelbash, 
{\tt hep-th/9606043}.}

\REF\DANIELSSON{U.H. Danielsson and B. Sundborg, \pl {\bf 358} (1995) 73 {\tt
 (hep-th/9504102)}.}

\REF\THEISEN{A. Klemm, W. Lerche and S. Theisen, \ijmp {\bf A11} (1996)
1929 {\tt (hep-th/9505150)}.}

\REF\SHENKER{M. Douglas and S.H. Shenker, \np {\bf B447} (1995) 271  
{\tt (hep-th/9503163)}.}

\REF\DOUGLAS{P.C. Argyres and M. Douglas, \np {\bf B448} (1995) 166
{\tt (hep-th/9505062)}.}

\REF\PLESSER{P.C. Argyres, M.R. Plesser and A.D. Shapere, \prl {\bf 75}
(1995) 1699 {\tt (hep-th/99505100)}.}

\REF\EGUCHI{T. Eguchi, K. Hori, K. Ito and S. -K. Yang 
\np {\bf B471} (1996) 430 {\tt (hep-th/9603002)}.}

\REF\SUNDBORG{U.H. Danielsson and B. Sundborg, \pl {\bf B370} (1996) 83
 {\tt (hep-th/9511180)}.}

\REF\SEIBERG{P.C. Argyres, M.R. Plesser and N. Seiberg, 
{\tt (hep-th/9603042)}.}

\REF\MINAHAN{J.A. Minahan and D. Nemeshansky, \np {\bf B464} 
(1996) 3 {\tt (hep-th/9507032)}.}

\REF\DHOKER{E. D'Hoker, I. Krichever, and D. Phong \np {\bf B489} (1997) 179
{\tt (hep-th/9609041)}; \np {\bf B489} (1997) 211 {\tt (hep-th/9609145)}; 
{\tt hep-th/9610156}; E. D'Hoker and D. Phong {\tt (hep-th/9701055)}.}

\REF\MATONE{M. Matone, Phys. Lett. {\bf B357} (1995) 342; 
G. Bonelli and M. Matone, \prl {\bf 76} (1996) 4107; 
G. Bonelli and M. Matone, ({\tt hep-th/9605090})
\pr{\bf 77} (1996) 4712;  
G. Bonelli, M. Matone and M. Tonin, ({\tt hep-th/9610026}).}

\REF\ALVAREZ{L. \'Alvarez-Gaum\'{e}, J. Distler, and M. Mari\~{n}o,
Int. J. Mod. Phys. {\bf A11} (1996) 4745 
({\tt hep-th/9604004});
L. \'Alvarez-Gaum\'{e} and M. Mari\~{n}o,
Int. J. Mod. Phys. {\bf A12} (1997) 975 
({\tt hep-th/9606191}); ({\tt hep-th/9606168});
L. \'Alvarez-Gaum\'{e} and S. F. Hassan, ({\tt hep-th/9701069});
L. \'Alvarez-Gaum\'{e}, M. Mari\~{n}o, and
F. Zamora ({\tt hep-th/9703072}); N. Evans, S.D.H. Hsu,
and M. Schwetz ({\tt hep-th/9608135}).}

\REF\ITO{K. Ito and S.-K. Yang, \pr {\bf D53} (1996) 2213 
{\tt (hep-th/9603073)};  \pl {\bf B366} (1996) 165 
{\tt (hep-th/9507144)}.}

\REF\SASAKURA{K. Ito and N. Sasakura, {\tt  (hep-th/9608054)}
\np {\bf B484} (1997) 141;
K. Ito, {\tt (hep-th/9703180)}.}

\REF\NOS{J.M. Isidro, A. Mukherjee, J.P. Nunes, and H.J. Schnitzer,
\np {\bf B} (to be published), {\tt hep-th/9609116}; {\tt hep-th/9703176}}

\REF\OHTA{Y. Ohta, {\tt hep-th/9604051, 9604059}.}

\REF\OZ{A. Hanany and Y. Oz\journal\np&B452(95)283 {\tt (hep-th /9505075)}.}

\REF\HANANY{A. Hanany\journal\np&B466(96)85 {\tt (hep-th/9509176)}.}
\REF\EGUCHI{T. Eguchi and K. Hori, {\tt hep-th/9607125}.}

\REF\RYANG{S. Ryang, \pl {\bf B365} (1996) 113  
{\tt (hep-th/9508163)}.}

\REF\BILAL{A. Bilal, {\tt (hep-th/9601007)}; 
A. Bilal and F. Ferrari, {\tt (hep-th/9605101)}
\np {\bf B480} (1996) 589;
A. Bilal and F. Ferrari, {\tt (hep-th/9602082)}
\np {\bf B469} (1996) 337; 
{\tt (hep-th/9606111)};
F. Ferrari, {\tt (hep-th/9702166)}; J. Schulze and N. P. Warner, 
{\tt (hep-th/9702012)}; J. M. Rabin, {\tt (hep-th/9703145)}.}

\REF\EWEN{H. Ewen, K. Foerger and S. Theisen,  
\np {\bf B485} (1997) 63
{\tt (hep-th/9609062)}; H. Ewen and K. Foerger, 
{\tt hep-th/9610049}.}

\REF\ALI{M. Alishahiha, {\tt hep-th/9609157}, {\tt hep-th/9703186}.}

\REF\SLATER{M. Slater, ({\tt hep-th/9601170}); N. Dorey, V.
Khoze, and M. Mattis, ({\tt hep-th/9612231});
\pr {\bf D54} (1996) 2921, ({\tt hep-th/9603136});
\pl {\bf B388} (1996) 324, ({\tt hep-th/9607066});
\pr {\bf D54} (1996) 7832, ({\tt hep-th/9607202});
\pl {\bf B390} (1997) 205, ({\tt hep-th/9606199});
K. Ito and N. Sasakura, \pl {\bf B382} (1996) 95,
({\tt hep-th/9602073}); ({\tt hep-th/9609104});
H. Aoyama, {\it et al.}, \pl {\bf B388} (1996) 331,
({\tt hep-th/9607076}); T. Harano and M. Sato,
({\tt hep-th/9608060}).}

\REF\FARKAS{H.M. Farkas and I. Kra, {\it Riemann Surfaces}, 2nd edn.,  
Graduate
 Texts in Mathematics vol. 71, Springer-Verlag, New York, 1991.}

\REF\MUKHERJEE{A. Mukherjee, {\it Aspects of Topological CFT}, PhD  Thesis,
Cambridge University, 1994.}

\break

\chap{Introduction.}

New insights into the behaviour of low energy supersymmetric gauge theories 
have led to an explosion
of activity in the subject [\SW].  One aspect which has attracted attention is 
the low-energy
properties of the Coulomb phase of $N$=2 gauge theories with $N_f$ matter 
multiplets in the fundamental representation, with and
without bare masses.  Since the moduli space of these gauge theories coincides
 with that of a
particular hyperelliptic curve [\SW]--[\MINAHAN], [\OZ, \HANANY], one attempts
 to extract the
strong-coupling physics from the curve describing the theory in question.  
The desired information
can be found from the prepotential
${\cal F} (\vec{a})$ characterising the low-energy effective Lagrangian, 
and from the 
Seiberg--Witten (SW) period integrals
$$ {\vec \Pi}=\pmatrix{{\vec a}_D\cr {\vec a}\cr}.
\eqn\zza
$$ These are related to ${\cal F} (\vec{a})$ by
$$ a_D^i={\p {\cal F}\o \p a_i}.
\eqn\zzb
$$ In the weak-coupling region one can write (schematically)
$$ {\cal F} (a) = {\cal F}_{{\hbox{\rm 1-loop}}} (a) + 
{\cal F}_{\hbox{\rm instantons}}(a).
\eqn\zzc
$$ A number of different methods have emerged for evaluating equation \zzc\ 
for a variety of groups,
either using the properties of hyperelliptic curves, or from the microscopic
 Lagrangian itself
 [\SLATER].  However, the available results for the strong-coupling regions 
are still rather sparse,
and as yet no complete picture has emerged, except for gauge groups of rank 1
 or 2.

One strategy for obtaining strong-coupling information is to derive and solve
 a set of Picard--Fuchs
(PF) equations for the SW period integrals.  The PF equations have been 
derived in a number of
special cases for $N_f = 0$ and massless multiplets for $N_f \neq 0$ 
[\THEISEN, \ITO, \SASAKURA,
\NOS, \ALI].

The PF equations have also been formulated [\SASAKURA, \NOS, \ALI] as an
explicit,
 analytic set of PF equations
valid for arbitrary classical gauge groups, and an extensive set of values
 for $N_f$ massless
multiplets in the fundamental representation. A systematic method for finding 
PF equations
for cases with bare mass $m=0$, which is particularly convenient for symbolic 
computer computations,
was given by Isidro {\it et al} [\NOS]. Alternative techniques, not involving 
PF equations, are
explored in [\DHOKER, \MATONE].

The situation for $m \neq 0$ is less fully developed [\SW, \ALVAREZ, \OHTA, 
\BILAL, \EWEN]. Part of
the problem is that the preferred SW differential $\lambda_{SW}$, is an 
abelian differential of the
third kind if
$m \neq 0$, while of the second-kind if $m=0$.

The SW differential has the property that
$$ 
a_i=\oint_{\gamma_i} \lambda_{SW},\qquad a^D_i=\oint_{\gamma^D_i}
\lambda_{SW},
\eqn\zzd
$$ 
where $\gamma_i$ and $\gamma_i^D$ are closed 1-cycles with canonical
intersection in a homology basis.  Since
$\lambda_{SW}$ is of the third-kind when
$m \neq 0$, {\it i.e.}, it has poles with non-vanishing residues, some care 
is required to
formulate the PF equations. Explicit PF equations for $SU(2)$ and $SU(3)$ have
 been presented for
$m\neq 0$ [\OHTA, \EWEN].  Even for these simplest examples several pages are 
required to write out
the actual PF equations.  Thus, it does not seem practical to present explicit 
PF equations for
arbitrary gauge groups and $N_f$ consistent with asymptotic freedom when
$m\neq 0$.  Rather, it seems more convenient to present a comprehensive 
set of explicit
algorithms, from which one may obtain the PF equations with
$m\neq 0$.

In this paper we generalise the results of [\NOS], and present a new method to
 obtain the PF
equations for arbitrary classical gauge groups, with massive matter 
hypermultiplets in the
fundamental representation. We have obtained the explicit PF equations for
 a number of gauge groups
with massive multiplets. The corresponding PF equations turn out to be 
extremely lengthy.
Therefore, there is no virtue in presenting these results. However, 
explicit results using our
methods coincide with PF equations previously presented in the literature
 for $m\neq 0$. Thus our
paper concentrates on a description of our method. In section 2
we formulate the problem, while in section 3 a necessary set of recursion 
relations are exhibited. 
Sections 4 and 5 describe how the PF equations are derived from this 
information.  Final  comments, as well as a summary and conclusions,
appear in sections 6 and 7, respectively.

\chap{Formulation of the problem.}

The strategy used in [\NOS] to derive the PF
equations of effective $N=2$ supersymmetric Yang--Mills theories in 4 
dimensions can be
modified to include massive matter hypermultiplets. To do so, let us
first  recall the necessary elements from [\NOS].

Let us consider an effective $N=2$ supersymmetric Yang--Mills theory
characterised by a
 certain gauge
group $G$ with rank $r$, a number $N_f$ of  massive matter hypermultiplets 
with bare masses
$m_j$, where $1\leq j\leq N_f$, and  a number of moduli $u_i$, where 
$1\leq i\leq r$. The matter
hypermultiplets will be taken in the fundamental representation, and 
$N_f$ will be restricted to
those values for which the theory is
asymptotically free, but for the moment $G$ will remain unspecified. 
Consider the complex algebraic
curves
$$
y^2=P(x;u_i;m_j;\Lambda),
\eqn\za
$$
where $P$ is a polynomial in $x$ of degree $2g+2$, and $\Lambda$ is  the 
dynamically
generated quantum scale. When all roots of $P$ are pairwise different, {\ie},
  away from
the zero locus of the discriminant of the curve, equation  \za\ defines a 
family of non-singular
hyperelliptic  Riemann  surfaces $\Sigma_g$ of genus $g$ [\FARKAS]. Under
an appropriate choice of the polynomial $P$,  the moduli space of quantum 
vacua of
the theory under consideration is coincident with that of the curves defined 
by
equation \za. On
$\Sigma_g$ there are $g$ holomorphic 1-forms
$$ 
x^j\,{\d x\o y},\qquad j=0, 1, \ldots, g-1.
\eqn\zb
$$ 
The following  $g$  1-forms  are meromorphic on $\Sigma_g$ and have 
pole singularities at infinity of order greater than 1\foot{The differential
$x^{g+1}{\rm d}x/y$ is of the second kind, \ie, it has vanishing residues,
for the particular curves of Seiberg-Witten theories. The differentials
$x^n{\rm d}x/y$  have non-vanishing residues at infinity when $n>g+1$. 
This corrects a mistake
in the terminology of [\NOS].}
$$
 x^j\,{\d x\o y},\qquad j=g+1, g+2, \ldots, 2g.
\eqn\zc
$$ 
{}Furthermore, the 1-form
$$
x^g{\d x\o y}
\eqn\zd
$$ 
is meromorphic on $\Sigma_g$, with a simple pole at infinity.
Altogether, the   abelian
differentials $x^j\d x/y$ in equations
\zb\ and \zc\ will be denoted collectively by $\omega_j$, where 
$j=0, 1, \ldots, 2g$, $j\neq g$.   We define the {\it basic range} $R$ to be
$R=\{0, 1, \ldots, \check g,\ldots 2g\}$, where a check over $g$ means the 
value
$g$ is to be omitted.

{}Following [\NOS] we give two definitions. Let us call
$W=y^2=P(x;u_i;m_j;\Lambda)$ in equation
\za. Moreover, given any differential $x^n\d x/y$, with $n\geq 0$   an
integer, let us define its {\it generalised $\mu$-period}
$\Omega_n^{(\mu)}(u_i;m_j;\Lambda;\gamma)$  along a fixed 1-cycle $\gamma\in
H_1(\Sigma_g)$  as the line integral [\MUKHERJEE]
$$
\Omega_n^{(\mu)}(u_i;m_j;\Lambda;\gamma):=(-1)^{\mu +1}\Gamma  (\mu + 1) 
\oint_{\gamma}{x^n\o W^{\mu + 1}}\,\d x.
\eqn\ze
$$ 
In equation \ze, $\Gamma$ stands for Euler's  gamma function, 
while $\gamma\in  H_1(\Sigma_g)$ is any  closed 1-cycle on the surface. 
For the
sake of simplicity, we will drop $u_i$, $m_j$, $\Lambda$ and $\gamma$ from the
notation for the periods $\Omega_n^{(\mu)}$. As explained in [\NOS],
we will work with an arbitrary value of $\mu$, which will only be set to 
$-1/2$
at the very end.

In effective 
$N=2$ supersymmetric gauge theories, there exists a preferred  differential,
 called the
{\it Seiberg--Witten (SW) differential}, $\lambda_{SW}$, with the following 
property [\SW]: the
electric and magnetic masses
$a_i$ and $a^D_i$ entering the BPS mass formula are given by the
 periods of $\lambda_{SW}$ along some specified closed cycles $\gamma_i,
\gamma^D_i \in H_1(\Sigma_g)$, as in equation \zzd.
In these theories, the polynomial $P(x;u_i;m_j;\Lambda)$ is of the special
form $P=p^2(x)-G(x)$, for certain $p(x)$ and $G(x)$ (see below). Then, the SW
differential takes on the following expression:
$$
\lambda_{SW}= {x\o y} \Big({G'\o G} {p\o 2} - p'\Big) \d x.
\eqn\nueva
$$
The SW differential further enjoys the property that its
modular   derivatives 
$\p\lambda_{SW}/\p u_i$ are (linear combinations of the)  holomorphic 
1-forms [\SW]. This ensures
positivity of the K\"ahler metric on moduli space. 

\chap{The recursion relations.}

As seen in the massless case treated in [\NOS], the periods 
$\Omega^{(\mu)}_n$
defined in equation \ze\ satisfy a set of recursion relations 
in both indices $n$ and $\mu$ that can
be used to derive the PF equations. Similar conclusions continue 
to hold for the massive
case as well. However, the polynomial $P(x;u_i;m_j;\Lambda)$ defining
 the curve depends on the
particular gauge group $G$ in such a way that a general expression valid 
for all
$G$, such as the one used in [\NOS], cannot be given. Instead, the different 
gauge groups have to
be treated separately. As the derivation of the recursion relations
follows the same pattern used in [\NOS] for the massless case, it will
not be reproduced here. We will simply list the final results below.
\item a) $G=SU(N_c)$, $N_f<N_c$.

The curve is given by [\OZ]
$$
W=p^2(x)-\Lambda_{N_f}^{2N_c-N_f} \prod_{j=1}^{N_f}(x+m_j),
\eqn\zf
$$
where
$$
p(x)=\sum_{i=0}^{N_c}u_ix^i,\qquad u_{N_c}=1,\qquad u_{N_c-1}=0.
\eqn\zfi
$$
Defining the symmetric polynomials in the masses $S_{N_f-j}(m)$ through the
expansion
$$
\prod_{j=1}^{N_f}(x+m_j)=\sum_{j=0}^{N_f}S_{N_f-j}(m)x^j,
\eqn\zg
$$
one finds that the following recursion relations hold:
$$
\eqalign{
&\Omega_{n+2N_c}^{(\mu+1)}=\Lambda_{N_f}^{2N_c-N_f}\sum_{j=0}^{N_f}S_{N_f-j}
\Omega_{n+j}^{(\mu+1)}-
(1+\mu)\Omega_n^{(\mu)}\cr
&-\sum_{i=0}^{N_c-1}\sum_{j=0}^{N_c-1}u_iu_j\Omega_{n+i+j}^{(\mu+1)}-2
\sum_{j=0}^{N_c-1}u_j
\Omega_{N_c+n+j}^{(\mu+1)}\cr}
\eqn\zba
$$
and
$$
\eqalign{
&\Omega_n^{(\mu)}= {-1\o
n+1-2N_c(1+\mu)}\Big[\Lambda_{N_f}^{2N_c-N_f}\sum_{j=0}^{N_f}
(2N_c-j)S_{N_f-j}\Omega_{n+j}^{(\mu+1)}\cr
&+\sum_{j=0}^{N_c-1}\sum_{l=0}^{N_c-1}(j+l-2N_c)u_ju_l\Omega_{n+j+l}^
{(\mu+1)}+
2\sum_{j=0}^{N_c-1}(j-N_c)u_j\Omega_{N_c+n+j}^{(\mu+1)}\Big].\cr}
\eqn\zh
$$
When $n+1-2N_c(2+\mu)\neq 0$, one can combine equations \zba\ and \zh\ 
to obtain, after shifting
$n+2N_c\rightarrow n$, 
$$
\eqalign{
&\Omega_{n}^{(\mu+1)}={1\o
n+1-2N_c(2+\mu)}\times\cr
&\Big[\Lambda_{N_f}^{2N_c-N_f}\sum_{j=0}^{N_f}
\big(n-2N_c+1-j(1+\mu)
\big)S_{N_f-j}\Omega_{n-2N_c+j}^{(\mu+1)}\cr
&+2\sum_{j=0}^{N_c-1}\big((1+\mu)(N_c+j)-(n-2N_c+1)\big)u_j\Omega_{n-N_c+j}^
{(\mu+1)}\cr
&+\sum_{j=0}^{N_c-1}\sum_{l=0}^{N_c-1}\big((j+l)(1+\mu)-(n-2N_c+1)\big)u_j
u_l\Omega_{n+j+l-2N_c}^{(\mu+1)}
\Big].\cr}
\eqn\zj
$$
Modular derivatives of periods are given by
$$
{\p\Omega_n^{(\mu)}\o\p u_i}=2\sum_{j=0}^{N_c}u_j\Omega_{n+i+j}^{(\mu+1)}.
\eqn\zk
$$

\item b) $G=SU(N_c)$, $2< N_c\leq N_f<2N_c$.

The curve is given by [\OZ]
$$
W=\Big[p(x)+{1 \o 4}\Lambda^{2N_c-N_f}_{N_f}\sum_{j=0}^{N_f-N_c}
S_{j}x^{N_f-N_c-j}\Big]^2-
\Lambda^{2N_c-N_f}_{N_f}\prod_{j=1}^{N_f}(x+m_j),
\eqn\zq
$$
with $p(x)$ as in equation \zfi. We observe that this curve can be obtained
 from the one given in
equations \zf\ and \zfi\ by a shift of some of the moduli $u_i$:
$$
u_i\rightarrow u_i+{1\o 4} \Lambda_{N_f}^{2N_c-N_f} S_{N_f-N_c-i}, \qquad 0\leq i\leq N_f-N_c.
\eqn\zqi
$$
In the classical limit, this redefinition does not affect the moduli. 
Applying this shift to
equations \zba\ through \zk\ we can straightforwardly derive the recursion 
relations corresponding
to this case from those of the previous case. Alternatively, 
one could carry out a step-by-step
derivation like the one needed  for $N_f<N_c$. Either way, the recursion
 relations turn out to be
given by
$$
\eqalign{
&\Omega_{n+2N_c}^{(\mu+1)}=\Lambda_{N_f}^{2N_c-N_f}\sum_{j=0}^{N_f}S_{N_f-j}
\Omega_{n+j}^{(\mu+1)}
-\sum_{j=0}^{N_c-1}\sum_{l=0}^{N_c-1}u_ju_l\Omega_{n+j+l}^{(\mu+1)}\cr
&-2\sum_{j=0}^{N_c-1}u_j
\Omega_{N_c+n+j}^{(\mu+1)}
-{1\o
2}\Lambda_{N_f}^{2N_c-N_f}\sum_{l=0}^{N_c}\sum_{i=0}^{N_f-N_c}u_lS_i
\Omega_{N_f-N_c+n-i+l}^{(\mu+1)}\cr
&-{1\o
16}\Lambda_{N_f}^{4N_c-2N_f}\sum_{j=0}^{N_f-N_c}\sum_{l=0}^{N_f-N_c}S_jS_l
\Omega_{2N_f-2N_c+n-j-l}^{(\mu+1)}-(1+\mu)\Omega_n^{(\mu)}\cr}
\eqn\zbc
$$
and
$$
\eqalign{  
&\Omega_n^{(\mu)}= {-1\o n+1-2N_c(1+\mu)}
\Big[\Lambda^{2N_c-N_f}_{N_f}\sum_{j=0}^{N_f}
(2N_c-j)S_{N_f-j}\Omega_{n+j}^{(\mu+1)}\cr
&+2\sum_{j=0}^{N_c-1}(j-N_c)u_j\Omega_{N_c+n+j}^{(\mu+1)}
+\sum_{j=0}^{N_c-1}\sum_{l=0}^{N_c-1}(j+l-2N_c)u_ju_l\Omega_{n+j+l}^
{(\mu+1)}\cr
&+{1\o
16}\Lambda^{4N_c-2N_f}_{N_f}\sum_{j=0}^{N_f-N_c}\sum_{l=0}^{N_f-N_c}
(2N_f-4N_c-j-l)S_{j}S_{l}\Omega^{(\mu+1)}_{n+2N_f-2N_c-j-l}\cr
&+{1 \o 2}\Lambda_{N_f}^{2N_c-N_f}\sum_{l=0}^{N_c}\sum_{j=0}^{N_f-N_c}
(N_f-3N_c-j+l)u_lS_{j}\Omega^{(\mu+1)}_{n+N_f-N_c-j+l}\Big].\cr}
\eqn\zr
$$
When $n+1-2N_c(2+\mu)\neq 0$, one can combine equations \zbc\ and \zr\ 
to obtain, after shifting
$n+2N_c\rightarrow n$,
$$
\eqalign{
&\Omega_n^{(\mu+1)}= {1\o n+1-2N_c(2+\mu)}
\Big[\Lambda^{2N_c-N_f}_{N_f}\sum_{j=0}^{N_f}\big(n-2N_c+1-j(1+\mu)\big)
S_{N_f-j}\Omega^{(\mu+1)}_{n-2N_c+j}\cr
&-{\Lambda_{N_f}^{4N_c-2N_f}\o 16}\sum_{l=0}^{N_f-N_c}\sum_{j=0}^{N_f-N_c}
\big(n-2N_c+1-(1+\mu)(2N_f-2N_c-j-l)\big)S_{j}S_{l}
\Omega^{(\mu+1)}_{n+2N_f-4N_c-j-l}\cr
&-{1\o 2}\Lambda_{N_f}^{2N_c-N_f}\sum_{l=0}^{N_c}\sum_{j=0}^{N_f-N_c}
\big(n-2N_c+1-(1+\mu)(l-j+N_f-N_c)\big)u_lS_{j}
\Omega^{(\mu+1)}_{n+N_f-3N_c-j+l}\cr
&-\sum_{j=0}^{N_c-1}\sum_{l=0}^{N_c-1}\big(n-2N_c+1-(1+\mu)(j+l)\big)
u_ju_l\Omega^{(\mu+1)}_{n-2N_c+j+l}\cr
&-2\sum_{j=0}^{N_c-1}\big(n-2N_c+1-(1+\mu)(j+N_c)\big)u_j
\Omega^{(\mu+1)}_{n-N_c+j}\Big].\cr}
\eqn\zs
$$
Modular derivatives of periods are given by
$$ {\p\Omega_n^{(\mu)}\o\p u_i}=2\sum_{j=0}^{N_c}u_j\Omega_{n+i+j}^{(\mu+1)}+
{1\o 2}\Lambda_{N_f}^{2N_c-N_f}\sum_{j=0}^{N_f-N_c}S_j
\Omega_{N_f-N_c+n-j+i}^{(\mu+1)}.
\eqn\zu
$$

\item c) $SO(N_c)$, $N_f<N_c-2$.

The curve for $N_f<N_c-r-2$ is given by [\HANANY]
$$
W=p^2(x)-\Lambda^{2(N_c-N_f-2)}_{N_f}x^d \prod_{j=1}^{N_f}(x^2-m_j^2),
\eqn\zbd
$$ 
with the rank $r$ and the power $d$ being respectively given by $r=N_c/2$ 
and $d=4$, if $N_c$ is
even, and $r=(N_c-1)/2$ and $d=2$, if $N_c$ is odd. The polynomial $p(x)$ 
is given by
$$
p(x)=\sum_{j=0}^{2r}u_jx^j, \qquad u_{2r}=1, \qquad u_{{\rm odd}}=0.
\eqn\zbe
$$
Let us expand the mass term of equation \zbd\ in terms of the symmetric 
polynomials in the {\it
squared}\/ masses $T_{N_f-j}(m^2)$ as follows:
$$
\prod_{j=1}^{N_f}(x^2-m_j^2)=\sum_{j=0}^{N_f}(-1)^{N_f-j}T_{N_f-j}(m^2)x^{2j}.
\eqn\zbf
$$
The above expansion could just as well be expressed as a double summation 
with coefficients given
by the symmetric polynomials in the masses $S_{N_f-j}(m)$, as done for 
$SU(N_c)$ in equation \zg.
However, we will find it more convenient to use the expansion \zbf, 
as it manifestly preserves the
even parity of the $SO(N_c)$ curve under $x\rightarrow -x$. Reasoning 
as in [\NOS], one finds that
the following recursion relations hold:
$$
\eqalign{
&\Omega_n^{(\mu)}={-1\o n+1 -4r(1+\mu)}\Big[
\Lambda_{N_f}^{2(N_c-N_f-2)}\sum_{i=0}^{N_f}(-1)^{N_f-i}(4r-2i-d)
T_{N_f-i}\Omega_{n+d+2i}^{(\mu+1)}\cr
&+\sum_{i=0}^{2r-2}\sum_{j=0}^{2r-2}(i+j-4r)u_iu_j
\Omega_{n+i+j}^{(\mu+1)}+2\sum_{j=0}^{2r-2}(j-2r)u_j\Omega_{n+j+2r}^{(\mu+1)}
\Big]\cr}
\eqn\zbg
$$
and 
$$
\eqalign{
&\Omega_n^{(\mu+1)}={1\o n+1-4r(2+\mu)}\times\cr
&\Big[\Lambda_{N_f}^{2(N_c-N_f-2)}\sum_{i=0}^{N_f}(-1)^{N_f-i}
\big(n-4r+1-(1+\mu)(2i+d)\big)
T_{N_f-i}\Omega_{n+d-4r+2i}^{(\mu+1)}\cr
&+2\sum_{j=0}^{2r-2}\big((1+\mu)(j+2r)-(n-4r+1)\big)u_j
\Omega_{n-2r+j}^{(\mu+1)}\cr
&+\sum_{i=0}^{2r-2}\sum_{j=0}^{2r-2}\big((1+\mu)(i+j)-(n-4r+1)\big)u_iu_j
\Omega_{n+i+j-4r}^{(\mu+1)}\Big].\cr}
\eqn\zbh
$$
Modular derivatives of periods are given by
$$
{\p\Omega_n^{(\mu)}\o \p u_i}=2\sum_{j=0}^{2r}u_j\Omega_{n+i+j}^{(\mu+1)}.
\eqn\zbi
$$

It is known from [\HANANY] that the curve for $N_c-r-2\leq N_f <N_c-2$
 can be obtained from that for
$N_f< N_c-r-2$ by a shift of the moduli similar to the one performed in 
equation \zqi. The
necessary shift now affects the even moduli only, and it involves the 
symmetric polynomials in the
squared masses $T_j(m^2)$ rather than the $S_j(m)$. Application of this 
shift to the recursion
relations given in equations
\zbg, \zbh\ and \zbi\ will produce the recursions corresponding to 
$N_c-r-2\leq N_f <N_c-2$.

\item d) $Sp(N_c)$, $0\leq N_f<N_c+2$.

The curve is given by [\SHAPERE, \EGUCHI]\foot{The sign of the $m^2$ term in
ref. [\EGUCHI] is opposite to that of eqn. (3.20). If so, that would mean the
double scaling limit would only apply to an even $N_f$. The replacement
$T_{N_f-j}(m^2)\rightarrow (-1)^{N_f-j}T_{N_f-j}(m^2)$ in eqns. (3.22), (3.23) 
and (3.24) below will accommodate the sign given in ref. [\EGUCHI].}
$$
x^2 W=p^2(x)-\Lambda_{N_f}^{2(2r+2-N_f)}\prod_{j=1}^{N_f}(x^2+m_j^2),
\eqn\zbl
$$
where $N_c=2r$. The polynomial $p(x)$ is given by
$$ 
p(x)=\sum_{j=2}^{2r+2}u_jx^j +\Lambda_{N_f}^{2r+2-N_f}\prod_{j=1}^{N_f}m_j,
\qquad u_{2r+2}=1, \qquad u_{{\rm odd}}=0.
\eqn\zbm
$$
The mass term of equation \zbl\ can be expanded in terms of the symmetric 
polynomials 
$T_{N_f-j}(m^2)$ defined by
$$
\prod_{j=1}^{N_f}(x^2+m_j^2)=\sum_{j=0}^{N_f}T_{N_f-j}(m^2)x^{2j}.
\eqn\zbmi
$$
Similar steps to the ones taken above then lead
to the following recursion relations:
$$
\eqalign{
&\Omega_{n}^{(\mu)}={-1\o
n+1-(4r+2)(1+\mu)}\Big[\Lambda_{N_f}^{2(2r+2-N_f)}
\sum_{j=1}^{N_f}\big(4(r+1)-2j\big)T_{N_f-j}
\Omega_{n+2j-2}^{(\mu+1)}\cr
&+2\sum_{i=2}^{2r}\big(i-2(r+1)\big)u_i\Omega_{n+2r+i}^{(\mu+1)}
+\sum_{i=2}^{2r}\sum_{j=2}^{2r}\big(i+j-4(r+1)\big)u_iu_j
\Omega_{n+i+j-2}^{(\mu+1)}\cr
&+2\Lambda_{N_f}^{2r+2-N_f}\prod_{i=1}^{N_f}m_j\sum_{j=2}^{2r+2}
\big(j-4(r+1)\big)u_j
\Omega_{n+j-2}^{(\mu+1)}\Big]\cr}
\eqn\zbn
$$
and
$$
\eqalign{
&\Omega_n^{(\mu+1)}={1\o n+1-(4r+2)(2+\mu)}\times\cr
&\Big[\sum_{i=2}^{2r}\sum_{j=2}^{2r}\big((i+j-2)(1+\mu)-(n-4r-1)\big)u_iu_j
\Omega_{n+i+j-4r-4}^{(\mu+1)}\cr
&+2\sum_{i=2}^{2r}\big((1+\mu)(2r+i)-(n-4r-1)\big)u_i
\Omega_{n-2r+i-2}^{(\mu+1)}\cr
&+2\Lambda_{N_f}^{2r+2-N_f}\prod_{i=1}^{N_f}m_j
\sum_{j=2}^{2r+2}\big((1+\mu)(j-2)-(n-4r-1)\big)u_j
\Omega_{n+j-4r-4}^{(\mu+1)}\cr
&+\Lambda_{N_f}^{2(2r+2-N_f)}\sum_{j=1}^{N_f}
\big((n-4r-1)+(1+\mu)(2-2j)\big)T_{N_f-j}
\Omega_{n+2j-4r-4}^{(\mu+1)}\Big].\cr}
\eqn\zbo
$$
Modular derivatives of periods are given by
$$ 
{\p\Omega_n^{(\mu)}\o \p u_i}=2\sum_{j=2}^{2r+2}u_j\Omega_{n+i+j-2}^{(\mu+1)}
+2\Lambda_{N_f}^{2r+2-N_f}\prod_{j=1}^{N_f}m_j \Omega_{n+i-2}^{(\mu+1)}.
\eqn\zbp
$$

\chap{Derivation of the Picard--Fuchs equations.}

{}Following the steps of [\NOS], one can use the recursion relations of 
the previous
section to derive a coupled system of first-order, partial differential 
equations (with respect to
the moduli) satisfied by the periods. We first set $\mu=-1/2$ in all 
what follows. Then we need to
identify the appropriate subspace of 
periods that one must restrict to, in order to properly solve the above 
recursions. We recall that
a key element is the behaviour of the curves under the operation of parity
$x\rightarrow -x$. A glance at the equations of the previous section 
immediately reveals that
conclusions completely analogous to those of [\NOS] continue 
to hold for the massive case. Let us
briefly recall them. 

$\bullet$ {}For the $SO(N_c)$ gauge groups, one must restrict to 
the {\it even}\/ subspace of the
basic range $R$, \ie, to even values of the subindex $n$. 
This follows from two facts. One is that
all the odd Casimirs of the gauge group $SO(N_c)$ vanish, 
so the recursion relations have a step of
2 units. Furthermore, the solution to those recursions can be expressed
 in terms of a set of initial
data with an even value for the subindex. As a function of the rank $r$, 
the genus $g$ of the
$SO(N_c)$ curve is $g=2r-1$. This being odd, the value of the subindex $n$ 
at which the recursion
\zbg\ blows up is skipped. Similarly, the zero of the denominator of equation
 \zbh\ is avoided when
$n$ is even.

$\bullet$ {}For the $Sp(N_c)$ gauge groups, one must restrict to the 
{\it odd}\/ subspace of the
basic range $R$, \ie, to odd values of the subindex $n$. Again, 
this follows from the same facts as
above. All the odd Casimirs of the gauge group $Sp(N_c)$ vanish, 
so the recursion relations have a
step of 2 units, but the solution to those recursions can be expressed 
in terms of a set of initial
data with an odd value for the subindex. As a function of the rank $r$, 
the genus $g$ of the
$Sp(N_c)$ curve is $g=2r$. This being even, the value of the subindex $n$ 
at which the recursion
\zbn\ blows up is skipped. Similarly, the zero of the denominator of 
equation \zbo\ is avoided when
$n$ is odd. The factor of $x^2$ present in the left-hand side of equation
\zbl\ is responsible for this odd parity, as opposed to the even parity 
of the $SO(N_c)$ recursions.

$\bullet$ {}For the $SU(N_c)$ gauge groups, the curve has no well defined 
parity under $x\rightarrow
-x$. This is a consequence of the fact that $SU(N_c)$ has both even and 
odd Casimirs which, in
turn, causes the recursion relations have a step of 1 unit. To solve the 
recursions,
one must first work on the enlarged subspace of periods  given by 
$R\cup\{g\}$,  \ie, 
$\Omega_n^{(\pm 1/2)}$, with $n\in R\cup\{g\}$. This is in order to avoid
 the divergence of the
recursion relations \zh\ and \zr\ that occurs when the subindex $n$ 
takes on the value $n=g$,
where the genus $g$ now equals the rank $r$, $g=r$. Next, one applies 
a linear relation satisfied by
the $\Omega_n^{(+1/2)}$, where $n \in R\cup\{g\}$. For a derivation 
of this linear relation that also
holds in the massive case treated here, see [\NOS].

Once the correct subspace of periods has been identified, 
the recursion relations
can be solved as explained in [\NOS]. Let us arrange the periods
 $\Omega_n^{(\pm 1/2)}$,
where $n$ spans the appropriate subspace of $R$, 
as  column
vectors: $\Omega^{(\pm 1/2)}=(\Omega_1^{(\pm 1/2)}, \ldots, 
\Omega_{r}^{(\pm 1/2)},\Omega_{r+1}^{(\pm 1/2)}, \ldots, 
\Omega_{2r}^{(\pm 1/2)})^t$. We have called
the dimension of the subspace under consideration $2r$, 
as it turns out to equal two times the rank
$r$  of the gauge group. We have also arranged the entries of 
$\Omega^{(\pm 1/2)}$ in
such a way that the first $r$ of them are holomorphic, while the last
 $r$ of them are meromorphic.
{}From the recursion relations, the vectors of periods $\Omega^{(-1/2)}$ 
and $\Omega^{(+1/2)}$ are
linearly related through a matrix
$M$
$$
\Omega^{(-1/2)}=M\cdot \Omega^{(+1/2)},
\eqn\zv
$$
where $M$ is  $(2r\times 2r)$-dimensional. Its entries  are certain polynomial 
functions
in the moduli $u_i$, the bare masses $m_j$ and the quantum scale $\Lambda$,  
explicitly computable
using the recursion relations given in the previous section.

Similarly, one can  use the expressions  for the modular derivatives 
(equations \zk, \zu, \zbi\
and \zbp), together with the recursions, in order to write a system of
 equations which, in matrix
form, reads
$$
{\p\o\p u_i}\Omega^{(-1/2)}=D(u_i)\cdot \Omega^{(+1/2)}.
\eqn\zw
$$
The matrix $D(u_i)$ is $(2r\times 2r)$-dimensional. Again,  
its entries are some polynomial
functions in the moduli $u_i$, the bare masses $m_j$ and the quantum scale
 $\Lambda$, explicitly
computable from the recursion relations. Assume for the moment that the 
matrix
$M$ of equation \zv\ is invertible. Combining the latter with equation \zw\ 
one ends up with
$$
{\p\o\p u_i}\Omega^{(-1/2)}=U_i\cdot \Omega^{(-1/2)},
\eqn\zx
$$
where we have defined the matrix $U_i$ as
$$
U_i=D(u_i)\cdot M^{-1}.
\eqn\zy
$$
Equation \zx\ is a coupled system of first-order, partial differential 
equations satisfied by the
periods $\Omega^{(-1/2)}$: {\it the first-order PF equations}. 
It expresses the modular derivatives
of the basic periods $\Omega^{(-1/2)}$ (of the subspace under
consideration) as certain linear combinations of the same periods
$\Omega^{(-1/2)}$. The coefficients entering those combinations 
(\ie, the entries of the $U_i$
matrices) are some rational functions of the moduli $u_i$, the masses $m_j$
 and the quantum scale
$\Lambda$, explicitly computable from the recursion relations.

To close this section, let us comment on the invertibility of the matrix $M$ 
of equation \zv.
Explicit evaluation in a wide class of examples shows ${\rm det}\, M$ 
to be a product of the
factors of the discriminant $\Delta(u_i, m_j, \Lambda)$ of the corresponding
 curve, possibly with
different multiplicities, and up to an overall  non-zero constant. Therefore, 
the $M$ matrix encodes
the singularity structure of the curve, and it is invertible except at the 
singularities of moduli
space, \ie, except on the zero locus of the discriminant $\Delta(u_i, m_j, 
\Lambda)$ of the curve.
This conclusion holds with some {\it caveat}\/ when the gauge group is 
$SU(N_c)$, since ${\rm det}\,
M$ may then pick up some additional zeroes. This possible new zero locus of 
${\rm det}\, M$ occurs
for the same reasons already explained in [\NOS] for the massless case. 
The inclusion of non-zero
bare masses
$m_j$ does not alter the arguments given in [\NOS] as far as
${\rm det}\, M$ is concerned, and they continue to hold in the massive case 
as well.

\chap{Decoupling the Picard--Fuchs equations.}

In  principle, integration of the system \zx\ yields the periods as functions 
of the moduli
$u_i$. The particular 1-cycle $\gamma\in H_1(\Sigma_g)$  being integrated over 
appears in the 
specific choice  of  boundary conditions that one  makes. In practice, however,
 the fact
that the system \zx\ is coupled makes it very difficult to solve. A possible 
strategy is to
concentrate on one particular subset of periods and try to obtain a reduced 
system of  equations 
satisfied by it, at the cost of increasing their order. In [\NOS] we have 
made use of the fact 
that one can
perform a change of basis that included $\Omega_{SW}$, the period of
$\lambda_{SW}$, as a basic vector. The decoupling of the resulting equations
 then followed from the
property that the modular derivatives of the SW differential $\lambda_{SW}$ 
are the holomorphic
differentials of the appropriate subspace within which the recursions are 
being solved  ---we call
this the {\it potential property of $\lambda_{SW}$}. However, a similar 
change
of basis is inconvenient now, because $\lambda_{SW}$ is of the third kind. Let
us see how this difficulty can be  circumvented.

Consider the  $U_i$ matrix in equation \zx\  and block-decompose it as 
$$ 
U_i=\pmatrix{A_i&B_i\cr C_i&D_i},
\eqn\zz
$$
where all four blocks $A_i$, $B_i$, $C_i$ and $D_i$ are $r\times r$.  
Next take the equations for
the derivatives of the holomorphic periods, $\p\Omega_n/\p u_i$ ,  
$1\leq n\leq r$, and solve them
for the meromorphic periods $\Omega_n$, $r\leq n\leq 2r$, in terms of the 
holomorphic ones and their
modular derivatives. That is, consider\foot{For notational simplicity we 
have dropped the
superscript $\mu =-1/2$, with the understanding that it has been fixed.}
$$
{\p\o\p u_i}\pmatrix{\Omega_1\cr\vdots\cr\Omega_r\cr}- A_i\pmatrix{\Omega_1
\cr\vdots\cr\Omega_r\cr}= B_i\pmatrix{\Omega_{r+1}\cr\vdots\cr\Omega_{2r}\cr}.
\eqn\zaa
$$ 
Solving equation \zaa\ for the meromorphic periods involves inverting the  
matrix $B_i$. Although
we lack a formal proof that $B_i$ is invertible, when the rank $r$  of the 
gauge group is greater
than 1, ${\rm det}\, B_i$ turns out to have two types of zeroes on moduli 
space. The first zero locus
contains a product of the factors of the discriminant $\Delta(u_i, m_j, 
\Lambda)$, possibly with
different multiplicities. The second zero locus is unrelated to the 
discriminant. Away
from these singularities, $B_i$ is invertible so,   from equation \zaa,
$$
\pmatrix{\Omega_{r+1}\cr\vdots\cr\Omega_{2r}\cr}= B_i^{-1}\cdot\Big({\p\o\p
u_i}-A_i\Big)\pmatrix{\Omega_1\cr\vdots\cr
\Omega_r\cr}.
\eqn\zab
$$
Next, substitute the meromorphic periods \zab\ into the last $r$ equations 
of \zx,
$$
{\p\o\p u_i}\pmatrix{\Omega_{r+1}\cr\vdots\cr\Omega_{2r}\cr}= C_i
\pmatrix{\Omega_{1}
\cr\vdots\cr\Omega_{r}\cr}+ D_i\pmatrix{\Omega_{r+1}\cr\vdots\cr
\Omega_{2r}\cr},
\eqn\zac
$$ 
and rearrange terms in order to obtain\foot{No summation over $i$ is implied 
here.}
$$
\eqalign{
&{\p^2\o\p u^2_i}\pmatrix{\Omega_1\cr\vdots\cr\Omega_r\cr}
-\bigg[{\p B_i\o\p u_i} B_i^{-1}+A_i+B_iD_iB_i^{-1}\bigg]
{\p\o\p u_i}\pmatrix{\Omega_1
\cr\vdots\cr\Omega_r\cr}\cr
&+\bigg[B_iD_iB_i^{-1}A_i-B_iC_i+{\p B_i\o\p u_i}B_i^{-1}A_i-{\p A_i\o\p u_i}
\bigg]
\pmatrix{\Omega_{1}\cr\vdots\cr\Omega_{r}\cr}=0.\cr}
\eqn\zad
$$ 
Equation \zad\ is a second-order coupled system, satisfied by the holomorphic 
periods on the
curve: {\it the second-order PF equations}. In order to decouple them, one now 
employs
the potential property of the SW differential $\lambda_{SW}$,  
$$
{\p\o\p u_i}\lambda_{SW}=\omega_i, \qquad i=1, \ldots, r,
\eqn\zae
$$
and from here one concludes an analogous property for the corresponding 
periods,
$$
{\p\o\p u_i}\Omega_{SW}=\Omega_i, \qquad i=1, \ldots, r.
\eqn\zaf
$$
The passage from equation \zae\ to equation \zaf\ is justified, even though in 
the presence of non-zero bare
masses $m_j$ the SW differential $\lambda_{SW}$ is of the third kind. 
The reason is that, from
[\SW], the residues of $\lambda_{SW}$ are known to be some multiples of 
the bare masses $m_j$, and
are therefore independent of the moduli $u_i$.

{}Finally,  substitution of equation \zaf\ into  \zad\ produces a 
{\it decoupled} system of
third-order, partial differential equations for the SW period $\Omega_{SW}$, 
$$
\eqalign{ 
&{\p^2\o\p u^2_i}\pmatrix{\p_1\Omega_{SW}\cr\vdots\cr\p_r\Omega_{SW}\cr} -
\bigg[{\p
B_i\o\p u_i} B_i^{-1}+A_i+B_iD_iB_i^{-1}\bigg] {\p\o\p u_i}
\pmatrix{\p_1\Omega_{SW}
\cr\vdots\cr\p_r\Omega_{SW}\cr}\cr &+\bigg[B_iD_iB_i^{-1}A_i-B_iC_i+
{\p B_i\o\p
u_i}B_i^{-1}A_i-{\p A_i\o\p u_i}\bigg]
\pmatrix{\p_1\Omega_{SW}\cr\vdots\cr\p_r\Omega_{SW}\cr}=0,\cr}
\eqn\zadi
$$ 
which are {\it the third-order PF equations} of the massive $N=2$ theory.
\chap{Final Comments.}

We observe that the recursion relations on which our method is founded can 
be derived without
taking recourse to the SW differential $\lambda_{SW}$ or its period 
$\Omega_{SW}$: all that is
required is a knowledge of the curve. One first derives a set of recursion 
relations. Next one
solves them, to obtain a first-order system \zx\ whose submatrices $A_i$, 
$B_i$, $C_i$ and $D_i$
contain all the relevant information.  The final step is the decoupling 
procedure.
Therefore, whatever limits we may want to take in the equations above must 
be taken at the level of
the recursion relations. If the latter enjoy the correct limiting properties, 
so will the PF
equations derived from them.

Two limits are worth taking. One is that in which all bare masses $m_j$ tend 
to zero:
$m_j\rightarrow 0$. This is the reduction to the massless case. The other one 
is the
integrating out of one massive quark, also called {\it double-scaling 
limit}\/:
sending the $j$-th mass to infinity, $m_j\rightarrow \infty$, and the 
quantum scale to
zero, $\Lambda_{N_f}^p\rightarrow 0$, while keeping $m_j\Lambda_{N_f}^p$ 
constant and setting it
equal to the new quantum scale $\Lambda_{N_f-1}^{p'}$,   ($p$ and
$p'$ being  the required powers to
which $\Lambda_{N_f}$ and $\Lambda_{N_f-1}$ are raised, in the presence of 
$N_f$ and $N_f-1$
flavours, respectively). This removes one flavour from the problem. 
One can easily check that all
the recursion relations given in section 3 tend to the corresponding 
recursions (either massless, or
with one massive quark less) in the appropriate limits. This is a trivial 
consequence of the fact
that the massive curves themselves are so constructed as to reproduce both 
limits correctly. For the
massless limit in particular, the recursion relations already derived in 
[\NOS] are also correctly
reproduced. 

However, the limit in which all quarks become massless is more intriguing, 
in the following sense:
the final equations for $\Omega_{SW}$ are
third-order in the massive case, while they are only second-order in the 
massless case. How does
this reduction in the order of the equations take place?  This point has 
been argued in [\OHTA], but
let us see how it can be recast in our language.

As already remarked, the system of first-order equations \zx\ not only 
enjoys the correct
limiting properties, but its derivation follows the same pattern in both 
the massive and the
massless case, since no use is made of the SW differential $\lambda_{SW}$ or
 its period
$\Omega_{SW}$. However,
the passage from the first-order equations \zx\ to the second-order equations 
\zad\ (or equations
(3.10) of [\NOS]) differs in the massive and the massless cases.  In both 
cases one
solves for the meromorphic periods in terms of the holomorphic periods and 
their modular 
derivatives. However, in the massless case one can immediately  apply the 
potential property of the
SW period $\Omega_{SW}$, while in the massive case one still substitutes all
 the
meromorphic periods into the remaining first-order equations. This
further step must be taken because, the SW differential now being of the third
 kind, one cannot
apply the potential property directly. This accounts for the increase
in the order of the final equations. 

The limit $m_j\rightarrow 0$ in which all masses vanish must therefore be 
taken in the 
recursion relations, or at most at the level of the first-order equations 
\zx, \ie, prior to the
decoupling of the equations. The reason is that the decoupling procedure 
does not commute with the
limit, as it proceeds differently in the two cases. 

\chap{Summary and Conclusions.}

In this paper we have extended a previously described derivation of the 
Picard--Fuchs equations
[\NOS], in order to include the case of massive matter hypermultiplets
in the fundamental representation. 
It is systematic, well
suited for symbolic computer computations, and holds for any classical gauge 
group. Our method is
based on a set of recursion relations satisfied by the period integrals that 
one can define on a
hyperelliptic Riemann surface.  We explicitly focused on the case of effective 
$N=2$ supersymmetric
Yang--Mills theories in 4 dimensions, where the relevant Riemann surfaces are 
such that 
their moduli space coincides with the moduli space of quantum vacua of the 
theory under
consideration. From this point of view, the Picard--Fuchs equations have 
proved to be an important
tool in probing the structure of moduli space and in computing the full 
prepotential, including
instantons. 

{}For gauge groups with rank 2 or greater, the Picard--Fuchs equations in 
the presence of massive
matter become intractable. The goal of explicitly computing the full quantum 
prepotential of
effective $N=2$ theories may well have to be accomplished using alternative 
techniques, such as the
ones put forward in [\DHOKER, \MATONE]. However, we believe the method 
presented here may 
find application elsewhere, as our derivation is not limited to these 
specific areas, and its
algebraic nature lends itself easily to different uses. 

\ack{It is once again a great pleasure to thank Manuel Isidro San Juan 
for technical advice on the
use of {\sl Mathematica}. Discussions with Steve Naculich and Henric Rhedin
 are also gratefully
acknowledged.}

\refout

\appendix
\centerline{Computer Calculation of the PF Equations}

Let us return to equation \zad, where we have observed that the entries of 
the blocks $A_i$, $B_i$,
$C_i$ and $D_i$ are certain rational functions of the moduli $u_i$, the bare 
masses
$m_j$, and the quantum scale $\Lambda$. This rational character stems from 
the inversion of the $M$
matrix in equation
\zy, where division by
${\rm det}\, M$ is required. As explained, ${\rm det}\, M$ is also a 
polynomial in the moduli and
the bare masses. Transferring ${\rm det}\, M$ to the left-hand side of 
equation \zx\ leaves us with
$$ {\rm det}\, M\,{\p\o\p u_i}\Omega^{(-1/2)}=\tilde U_i\cdot 
\Omega^{(-1/2)},
\eqn\xa
$$ where $\tilde U_i$ is the matrix product $D(u_i)\cdot \tilde M ^{-1}$, 
$\tilde M$ being the matrix of
cofactors of $M$. Given that both $D(u_i)$ and $\tilde M$ have 
polynomial entries in the moduli and
the masses, $\tilde U_i$ also has purely polynomial entries, 
rather than rational functions.\foot{The
one possible exception to this polynomial character of $\tilde U_i$ 
may (but need not) occur for the
$SU(N_c)$ gauge groups, under the circumstances explained in [\NOS].} 
To the effect of performing
symbolic computer calculations, this has an obvious computational advantage, 
at the small expense of
having to modify equations \zx\ through
\zad. We first block-decompose $\tilde U_i$ in equation \xa\ as 
$$  
\tilde U_i=\pmatrix{\tilde A_i&\tilde B_i\cr \tilde C_i&\tilde D_i}.
\eqn\xb
$$ Next we solve for the meromorphic periods in terms of the holomorphic ones,
as in equation
\zab,
$$
\pmatrix{\Omega_{r+1}\cr\vdots\cr\Omega_{2r}\cr}= ({\rm det}\, 
M )\tilde B_i^{-1}\cdot {\p\o\p
u_i}\pmatrix{\Omega_1\cr\vdots\cr
\Omega_r\cr}-\tilde B_i^{-1} \tilde A_i\pmatrix{\Omega_1\cr\vdots\cr
\Omega_r\cr},
\eqn\xc
$$ and substitute into the remaining equations, as done in \zac. 
After some rearrangements we
find\foot{No summation over $i$ is implied here.}
$$
\eqalign{ &{\p^2\o\p u^2_i}\pmatrix{\Omega_1\cr\vdots\cr\Omega_r\cr} -
\bigg[{\p \tilde B_i\o\p u_i}
\tilde B_i^{-1} +({\rm det}\, M)^{-1}\tilde A_i\cr &+({\rm det}\, M)^{-1}
\tilde B_i\tilde D_i\tilde
B_i^{-1}-({\rm det}\, M)^{-1}{\p\o\p u_i}{\rm det}\, M\bigg] {\p\o\p u_i}
\pmatrix{\Omega_1
\cr\vdots\cr\Omega_r\cr}\cr &+\bigg[({\rm det}\, M)^{-2}\tilde B_i\tilde D_i
\tilde B_i^{-1}\tilde
A_i-({\rm det}\, M)^{-2} \tilde B_i\tilde C_i\cr &+({\rm det}\, M)^{-1}{\p
\tilde B_i\o\p u_i}\tilde B_i^{-1}\tilde A_i-({\rm det}\, M)^{-1}
{\p \tilde A_i\o\p u_i}\bigg]
\pmatrix{\Omega_{1}\cr\vdots\cr\Omega_{r}\cr}=0.\cr}
\eqn\xd
$$  Obviously, equation \xd\ reduces to \zad\ upon formally setting 
${\rm det}\, M=1$ and
eliminating the tildes, as this corresponds to leaving ${\rm det}\, M$ 
in the denominator in the
right-hand side of equation \xa. 

Below we list a {\sl Mathematica} programme that computes the matrices 
$\tilde A_i$, $\tilde B_i$, $\tilde C_i$ and $\tilde D_i$, as well as 
${\rm det}\, M$,  for the
$SO(N_c)$ gauge theory with $N_f$ massive multiplets, when $N_f<N_c-r-2$. 
It can be easily modified to
apply to the
$SU(N_c)$ and $Sp(N_c)$ gauge groups. For typographical reasons, 
the names of some variables
in the body of the programme have been changed with respect to the paper.
 Thus, $P_n$ stands
for $\Omega_n^{(+1/2)}$, $Q_n$ for
$\Omega_n^{(-1/2)}$, ``mu" for $\mu$, ``nf" for $N_f$, ``nc" for $N_c$, $R$ 
for
$\Delta(u_i, m_j, \Lambda)$, $TU[i]$ for $\tilde U_i$,  $TA[i]$ for 
$\tilde A_i$, 
$TB[i]$ for $\tilde B_i$, $TC[i]$ for $\tilde C_i$, and $TD[i]$ for 
$\tilde D_i$.
The names of other variables used are explained in the programme.
All the required equations are taken from the body of the paper.

{\tt

(* Input the data

	nf=number of flavours

	nc=number of colours

	d$=4$ if nc is even, d$=2$ if nc is odd
 
	r=rank=nc/2 if nc is even, r=rank=(nc-1)/2 if nc is odd
	
	mu$=-$1/2

	m[1], m[2], ..., m[nf]: bare masses of the multiplets

	T[0], T[1], ..., T[nf]: symmetric polynomials in the squared masses

	u[0], u[2], ..., u[2r-2]: moduli

and run *)

(* Known moduli for SO(nc) are the highest one, which is 1, 
and all the odd ones, which vanish *)

Do[u[j]=0, $\{$j, 1, 2r-1, 2$\}$]

u[2r]=1;

(* Definition of the curve W. For simplicity the quantum scale has been 
set to 1 *)

p[x]=Sum[u[j] x $\wedge$  j, $\{$j, 0, 2r$\}$];

W[x]=Collect[Expand[p[x]$\wedge$2-x$\wedge$d 
Product[(x$\wedge$2-m[j]$\wedge$2), $\{$j, 1, nf,
1$\}$]], x];

Print["p[x] = ", p[x]];

Print["W[x] = ", W[x]];

(* Discriminant of the curve *)

DxW=Simplify[D[W[x], x]];

R=Factor[Simplify[Resultant[W[x], DxW, x]]];

Print["R = ", R]

(* Definition of the list of initial values IV[j] that will close the 
recursions *)

Do[P[j]=IV[j], $\{$j, 0, 4r-2, 2$\}$]

initials=Table[IV[j], $\{$j, 0, 4r-2, 2$\}$];

Print["Initial values = ", initials]

(* Recursion relation for the P[j] *)

P[n$\_$Integer]=1/(n+1-4r(2+mu)) (Sum[(-1)$\wedge$(nf-i) 
(n-4r+1-(1+mu)(2i+d)) T[nf-i] P[n+d-4r+2i],
$\{$i, 0, nf, 1$\}$]+ 2Sum[((1+mu)(j+2r)-(n-4r+1)) u[j] P[n-2r+j], 
$\{$j, 0, 2r-2, 1$\}$]+
Sum[((1+mu)(i+j)-(n-4r+1)) u[i] u[j] P[n+i+j-4r], $\{$i, 0, 2r-2, 1$\}$,
 $\{$j, 0, 2r-2, 1$\}$]);

(* The required values for the P[j] are computed first for the sake of 
efficiency, then expressed in
terms of the initial values, and finally stored as the new variables PP[j] *)

Do[PP[j]=Collect[Simplify[P[j]], initials], $\{$j, 0, 8r-4, 2$\}$]

(* Recursion relation for the Q[j] in terms of the PP[j] *)

Q[n$\_$Integer]=-1/(n+1-4r(1+mu)) (Sum[(i+j-4r) u[i] u[j] PP[n+i+j], 
$\{$i, 0, 2r-2, 1$\}$, $\{$j,
0, 2r-2, 1$\}$]+ Sum[(-1)$\wedge$(nf-i) (4r-2i-d) T[nf-i] PP[n+d+2i], 
$\{$i, 0, nf, 1$\}$]+
2Sum[(j-2r) u[j] PP[n+j+2r], $\{$j, 0, 2r-2, 1$\}$]);

(* Values of P[j] or Q[j]  can always be expressed as linear combinations of 
the above initial
values,  with coefficients that are polynomials in the moduli *)

(* Computation of the M matrix relating the Q[j] to the P[j] *)

Do[QQ[j]=Collect[Simplify[Q[j]], initials], $\{$j, 0, 4r-2, 2$\}$]

M=Table[Coefficient[QQ[i], IV[j]], $\{$i, 0, 4r-2, 2$\}$, $\{$j, 0, 4r-2, 2$\}
$];

(* Next we check that the determinant of the M matrix, DM, contains the same 
factors as the
discriminant R of the curve, possibly with different multiplicities *)

DM=Factor[Det[M]];

Print["DM = ", DM]

(* Definition of modular derivatives of the Q[j]  in terms of the P[j] *)

DQ[n$\_$Integer, j$\_$Integer]=2Sum[u[i] PP[n+j+i], $\{$i, 0, 2r, 1$\}$];

(* Computation of the matrices DD[j] expressing modular derivatives of 
the Q[j] as linear
combinations of the P[j] *)

Do[
	
	Do[
		DQQ[n, j]=Collect[Simplify[DQ[n, j]], initials],
	$\{$n, 0, 4r-2, 2$\}$];
	
	DD[j]=Table[Coefficient[DQQ[n, j], IV[i]], $\{$n, 0, 4r-2, 2$\}$, 
$\{$i, 0, 4r-2, 2$\}$],
	
$\{$j, 0, 2r-2, 2$\}$]

(* Inversion of M.  We pull its determinant, DM, to  the left-hand side 
of the equations. SIM is the
matrix of cofactors of M *)

SIM=Simplify[DM*Inverse[M]];

(* Computation of the matrices TU[j] of coefficients in the expansion of 
the modular derivatives of
the Q[j] as linear combinations of  the Q[j] *)

Do[TU[j]=Simplify[DD[j].SIM], $\{$j, 0, 2r-2, 2$\}$]

(* Partition of the TU[j] into  submatrices: TA[j], TB[j], TC[j], TD[j] *)

Do[

	TA[j]=TU[j][[Range[1, r], Range[1, r]]];

	TB[j]=TU[j][[Range[1, r], Range[r+1, 2r]]];

	TC[j]=TU[j][[Range[r+1, 2r], Range[1, r]]];

	TD[j]=TU[j][[Range[r+1, 2r], Range[r+1, 2r]]],

$\{$j, 0, 2r-2, 2$\}$]	
	
Do[

	Print["----- Below is modulus ", u[j], " -----"];

	Print["TA[", j,"] = ", TA[j]];

	Print[""];

	Print["TB[", j,"] = ", TB[j]];

	Print[""];

	Print["Det[TB[", j, "]] = ", Factor[Det[TB[j]]]];

	Print[""];

	Print["TC[", j,"] = ", TC[j]];

	Print[""];

	Print["TD[", j,"] = ", TD[j]];

	Print[""],

$\{$j, 0, 2r-2, 2$\}$]

(* One can check that the determinants of the TB[j] contain the factors 
present in the 
discriminant R of the curve. *)}

\end